\documentclass[twocolumn,floatfix,pra,showpacs,preprintnumbers,superscriptaddress,amsmath,amssymb,10pt,aps]{revtex4}
\usepackage[toc,page]{appendix}
\usepackage{mathtools}
\usepackage{mathptmx}
\usepackage{subfigure}
\usepackage{psfrag,graphicx}
\usepackage{dcolumn}
\usepackage{amsmath,amssymb}
\usepackage{bm}
\usepackage{color}
\usepackage{latexsym}
\usepackage{epstopdf}
\usepackage{color}
\usepackage[english]{babel}
\usepackage{latexsym}
\usepackage{psfrag,graphicx}
\usepackage{subfigure}
\usepackage{amsmath}
\usepackage{amssymb}
\usepackage{amsfonts}
\usepackage{bm}
\usepackage{natbib}
\usepackage{epstopdf}
\usepackage{xcolor}
\usepackage{changes}
\usepackage{appendix}

\definecolor{mygrey}{gray}{0.35}
\definecolor{myblue}{rgb}{0.2,0.2,0.8}
\definecolor{myzard}{cmyk}{0,0,0.05,0}
\definecolor{mywhite}{rgb}{1,1,1}
\definecolor{mywhite}{rgb}{1,1,1}
\definecolor{myred}{rgb}{1,0.,0.3}

\usepackage[colorlinks=true,citecolor=myblue,linkcolor=myred]{hyperref}

\def\ba{\begin{align}}
	\def\enda{\end{align}}
\def\bi{\begin{itemize}}
	\def\ei{\end{itemize}}

\def\be{\begin{equation}}
	\def\ee{\end{equation}}
\def\bea{\begin{eqnarray}}
	\def\eea{\end{eqnarray}}
\def\bse{\begin{subequations}}
	\def\ese{\end{subequations}}

\definechangesauthor[name=moritz, color=pruple]{1}

\begin{document}
	\title{Mechanical Squeezed Kerr Oscillator based on Tapered Ion Trap }
	
	\author{Bogomila S. Nikolova}
	\affiliation{Center for Quantum Technologies, Department of Physics, St. Kliment Ohridski University of Sofia, James Bourchier 5 blvd, 1164 Sofia, Bulgaria}
	\author{Moritz G\"ob}
	\affiliation{Experimental Physics I, University of Kassel, Heinrich-Plett-Strasse 40, 34132 Kassel, Germany}
	\author{Kilian Singer}
	\affiliation{Experimental Physics I, University of Kassel, Heinrich-Plett-Strasse 40, 34132 Kassel, Germany}
	\author{Peter A. Ivanov}
	\affiliation{Center for Quantum Technologies, Department of Physics, St. Kliment Ohridski University of Sofia, James Bourchier 5 blvd, 1164 Sofia, Bulgaria}
	
	\begin{abstract}
		We propose the realization of a mechanically squeezed Kerr oscillator with a single ion in a tapered trap. We show that the motion coupling between the axial and radial modes caused by the trap geometry leads to Kerr nonlinearity of the radial mode with magnitude controlled by the trap frequencies. This allows the realization of non-Gaussian quantum gates, which play a significant role in the universal set of continuous variable quantum gates. Furthermore, we show that, because of the nonlinearity of the ion trap, applying an off-resonant time-varying electric field along the trap axis causes a motion squeezing of the radial mode. Finally, we discuss the motion mode frequency spectrum of an ion crystal in a tapered trap. We show that the frequency gap between the motion modes increases with trap nonlinearity, which benefits the realization of faster quantum gates.
	\end{abstract}

	\maketitle
	
	\section{Introduction}
	The creation and study of non-Gaussian state of motion of a quantum harmonic oscillator has attracted a great interest due to the wide range of applications including continuous variable (CV) quantum computing \cite{Braunstein2005,Chamberland2022}, bosonic quantum simulation \cite{Georgescu2014,Chen2023}, and quantum metrology \cite{Degen2017,Fadel2024}. Mechanical modes are fundamental building blocks for CV information processing, providing a large Hilbert space. The universal set of CV quantum gates includes Gaussian gates composed of displacement, single-mode squeezing, phase shift, and non-Gaussian gates such as Kerr nonlinearity- based phase gates. Nonlinear quantum evolution is difficult to realize since it usually requires off-resonance coupling between mechanical modes or between oscillator and a two-level system. Recently, the realization of Kerr nonlinearities has been demonstrated in superconducting circuits \cite{Marti2024,Iyama2024,Grimm2020,He2023} and trapped ions \cite{Roos2007,Ding2017,Ding2018}.
	
	In this paper, we exploit the effects of nonlinearities that arise from the funnel-shaped three-dimensional trapping potential experienced by the trapped ions. The tapered geometry of the ion trap leads to position-dependent radial trapping frequencies which increase along the trap axis. This type of trap geometry leads to a coupling of the motional radial and axial degrees of freedom of the trapped ions and to the realization of a Duffing-type nonlinear oscillator. The tapered geometry of the trap has recently been utilised for precise measurements of zeptonewton-scale forces, \cite{Kilian Zepto}, as well as for the experimental implementation of a single-atom heat engine \cite{Kilian Heat}. Driven nonlinear oscillators can be useful for characterization of a wide variety of phenomena since they experience squeezing, \cite{Yang,Huber}, phase transitions \cite{Wang}, chaos \cite{Goto2021}, with applications in quantum thermodynamics, \cite{Pijn,Kranzl}, quantum sensing, \cite{Hempel,Wolf,Milne}, quantum information \cite{Bruzewicz}, and quantum simulation \cite{Monroe}.
	
	Here we show that the single ion in the funnel-shaped trap can be used to realize a mechanical squeezed Kerr oscillator. Furthermore, we demonstrate that the application of a rapidly oscillating electric field along the axial direction ($z$-axis) leads to a suppression of phonon exchange between the axial and radial motional modes. The application of such an off-resonance force leads to negligibly small displacement experienced by the ion trapped in the conventional Paul trap. In contrast here such a field causes squeezing of the radial motion mode with an amplitude proportional to the magnitude of the applied electric field. Furthermore, we show that the coupling between the axial and radial motional modes of a single trapped ion causes a Kerr nonlinearity of the radial mode as well as a cross-Kerr nonlinearity between the two modes with magnitude controlled by the trap frequencies. This effect is beneficial for creation of non-Gaussian quantum gates \cite{Ricardo2024} as well as for high-precision quantum metrology \cite{Guo2024}. We also propose the implementation of a spin-dependent version of the mechanically squeezed Kerr oscillator, incorporating a spin-dependent force acting on the quantum oscillator, which, due to the trap-geometry-induced nonlinearity, creates a spin-dependent squeezing. We show that such interactions are useful and very effective in realization of non-Gaussian Schr\"odinger cat states.
	
	Finally, we extend the discussion to an ion crystal confined in a funnel-shaped trap. We show that the ion's equilibrium positions are not affected by the trap geometry. However, the tapered trap geometry breaks the reflection symmetry of the normal modes, compared to the conventional linear Paul trap. We show that the frequency gap between the radial normal modes increases with the parameter describing the length of the funnel. The realization of faster quantum gates benefits from the larger frequency gap, as it suppresses unwanted phonon excitations \cite{Zhu2006}.

	The paper is organized as follows. In Section \ref{NLO} we discuss the realization of a mechanically squeezed Kerr oscillator with ion in a funnel-shaped trap. The motional coupling between the axial and radial modes causes Kerr nonlinearity of the radial mode with magnitude controlled by the trap frequencies. We show that radial motion squeezing can be created using an oscillating electric field applied in axial direction. We also discuss the effect of spin-dependent laser forces applied on a single ion in a tapered trap. In Sect. \ref{CSKS} the realization of a non-Gaussian state of motion is discussed using Kerr nonlinearity, which can be used for high-precision quantum metrology. In Sect. \ref{IS} we discuss an ion crystal in a tapered trap. We show that the frequency gap between the motion modes increases with a larger angle of the funnel. Finally, the summary and the outlook are presented in Sect. \ref{SO}\\
	
	\begin{figure}
		\includegraphics[width=0.45\textwidth]{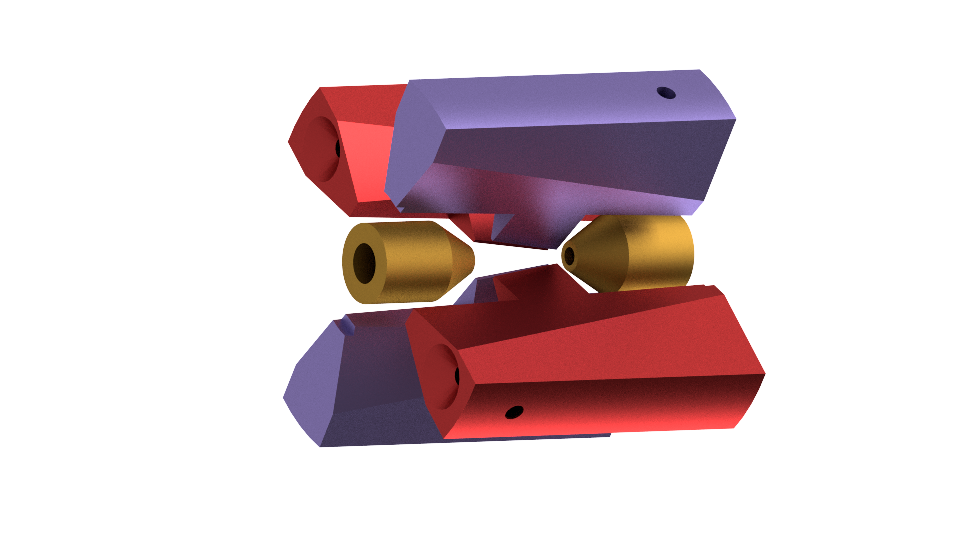}
		\caption{Computer-aided design of the tapered ion trap. The blue and red blade
 electrodes are supplied with bipolar radio-frequency drive. The bronce endcap
 electrodes are designed for allowing optical access along the axial $z$
 direction.}
		\label{fig11}
	\end{figure}

	\section{Non-linear quantum oscillator}\label{NLO}
	\subsection{Funnel-shaped potential}
	We consider a single ion with mass $m$ and charge $e$ confined in a funnel-shaped potential in which we observe stronger radial confinement with increasing axial position $z$. The radial trap frequencies can be well described by the relation $\omega_{x,y}(1-z/l_{0})^{-2}$, where $l_{0}$ is a parameter describing the strength of the funnel, see Fig. \ref{fig11}. In the tapered trap described in \cite{Deng2024} the electrodes have an angle with respect to the trap axis of 10$^\circ$ resulting in $l_0=1.81$mm. This value can be decreased by tilting the electrodes at a larger angle and by decreasing the distance between the radio frequency electrodes. Stable trapping of a linear ion crystal is achieved in this trap for axial trapping frequencies ranging between 100 kHz$/2\pi$ and 450 kHz$/2\pi$ and radial trapping frequencies typically are ranging between a $700$kHz$/2\pi$ and 1.8 MHz$/2\pi$. We assume that the radial trap frequency $\omega_{y}$ is sufficiently different from the other two trap frequencies such that the $y-$mode is not driven, and for simplicity we set $y=0$. In first order $|z/l_{0}|\ll 1$ the Hamiltonian for the two-dimensional $x$-$z$ nonlinear quantum oscillation dynamics is given by
	\begin{equation}
		\hat{H}_{\rm vib}=\frac{\hat{p}^{2}_{x}}{2m}+\frac{\hat{p}^{2}_{z}}{2m}+\frac{1}{2}m\omega^{2}_{x}\hat{r}_{x}^{2}+\frac{1}{2}m\omega^{2}_{z}\hat{r}_{z}^{2}+\frac{m\omega^{2}_{x}}{l_{0}}\hat{r}_{z}\hat{r}_{x}^{2},\label{H}
	\end{equation}
	where $\hat{r}_{k}$ and $\hat{p}_{k}$ ($k=x,z$) are the ion's position and momentum operators. The first four terms in (\ref{H}) describe the standard two-dimensional quantum harmonic oscillator, while the last term refers to the nonlinear coupling between motion along $x$- and $z$-directions.
	In the following, it is convenient to express these operators as $\hat{r}_{k}=r_{0k}(\hat{a}^{\dag}_{k}+\hat{a}_{k})$ and $\hat{p}_{k}=ip_{0k}(\hat{a}^{\dag}_{k}-\hat{a}_{k})$ where $r_{0k}=\sqrt{\hbar/2m\omega_{k}}$ and $\hat{p}_{0k}=\sqrt{\hbar m\omega_{k}/2}$, with $\hat{a}_{k}$ and $\hat{a}^{\dag}_{k}$ being the annihilation and creation operators of phonon excitation with frequency $\omega_{k}$ in mode $k$.
	
	\subsection{Time-varying electric field}
	\begin{figure}
		\includegraphics[width=0.45\textwidth]{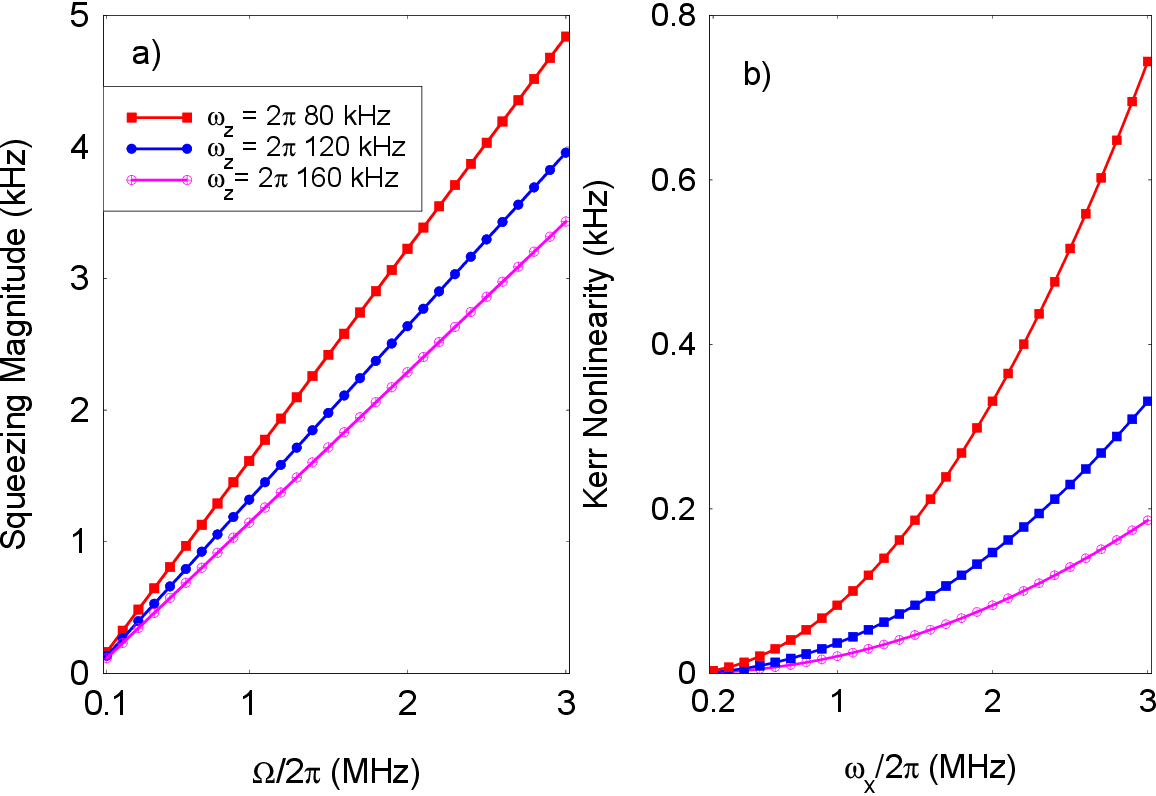}
		\caption{(a) Squeezing magnitude $\epsilon$ as a function of $\Omega$ for various $\omega_{z}$. The driving frequency is $\Phi/2\pi=2.5$ MHz. (b) Kerr nonlinearity as a function of the radial trap frequency $\omega_{x}$ for $l_{0}=0.05$ mm.}
		\label{fig1}
	\end{figure}
	
	We assume that a classical oscillation force along the axial direction is applied to the ion. The action of the force is displacing a motional amplitude of the axial oscillator and is described by the Hamiltonian
	\begin{equation}
		\hat{H}_{F}(t)=F_{z} r_{0z}\sin(\Phi t)(\hat{a}^{\dag}_{z}e^{i\phi}+\hat{a}_{z}e^{-i\phi}),\label{force}
	\end{equation}
	with force amplitude $F_{z}$. Here $\Phi$ is the frequency of the periodic driving and $\phi$ is the phase which we set $\phi=\pi/2$. In this work, we consider a driving frequency $\Phi\gg\omega_{z}$. The effect of such a time-varying electric field on the ion's motion is negligible, as long as the nonlinear coupling is not present. In the following, we will demonstrate that the nonlinear motion coupling leads to squeezing of the radial $x$-motion if combined with an off-resonant electric field along the $z$-axis.
	\subsection{Effective Hamiltonian}
	Including the time-varying electric field term (\ref{force}) the Hamiltonian becomes
	\begin{eqnarray}
		\hat{H}(t)&=&\hbar\omega_{x}\hat{a}^{\dag}_{x}\hat{a}_{x}+\hbar\omega_{z}\hat{a}^{\dag}_{z}\hat{a}_{z}+\hbar\lambda(\hat{a}^{\dag}_{z}+\hat{a}_{z})(\hat{a}^{\dag}_{x}+\hat{a}_{x})^{2}\notag\\
		&&-i\hbar\sin(\Phi t)\Omega(\hat{a}_{z}^{\dag}-\hat{a}_{z}),\label{H1}
	\end{eqnarray}
	where $\lambda=r_{0z}\omega_{x}/2 l_{0}$ is the nonlinear coupling, $\Omega=r_{0z}F_{z}/\hbar$ is the driving frequency of the off-resonant electric field along the $z$-axis. We have set the energy zero to eliminate  the zero- point energy terms. It is convenient to transform the Hamiltonian into interaction picture with respect to $\hat{H}_{0}=\hbar\omega_{x}\hat{n}_{x}+\hbar\omega_{z}\hat{n}_{z}$ ($\hat{n}_{k}=\hat{a}^{\dag}_{k}\hat{a}_{k}$ being the phonon number operator) such that we have
	\begin{eqnarray}
		\hat{H}_{R}(t)&=&\hbar\lambda\{(e^{i(2\omega_{x}+\omega_{z})t}\hat{a}^{\dag 2}_{x}\hat{a}^{\dag}_{z}+{\rm h.c.})+(e^{i(2\omega_{x}-\omega_{z})t}\hat{a}^{\dag 2}_{x}\hat{a}_{z}\notag\\
		&&+{\rm h.c.})\}+\hbar\lambda(\hat{a}^{\dag}_{z}e^{i\omega_{z}t}+\hat{a}_{z}e^{-i\omega_{z}t})(2\hat{n}_{x}+1)\label{Hr}\\
		&&+\frac{\hbar\Omega}{2}\{(e^{i(\Phi+\omega_{z})t}\hat{a}^{\dag}_{z}+{\rm h.c.})-(e^{i(\Phi-\omega_{z})t}\hat{a}_{z}+{\rm h.c.})\}.\notag
	\end{eqnarray}
	The first two terms in (\ref{Hr}) describe coherent phonon exchange between $x$- and $z$-vibrational modes in which creation of an axial phonon is accompanied by the creation of two radial phonons and, respectively, a process in which one axial phonon is converted into two radial phonons and vice versa. The third and fourth terms describe a coherent axial displacement that may depend on the radial phonon number. As long as the conditions $(2\omega_{x}\pm\omega_{z})\gg\lambda$, $\omega_{z}\gg\lambda$ and $|\Phi\pm\omega_{z}|\gg\Omega/2$ are fulfilled, one can perform time averaging of the rapidly oscillating terms in (\ref{Hr}). For arbitrary off-resonant driving frequency $\Phi$, the time-averaged dynamics generates a Kerr nonlinearity in the radial phonon mode. Crucially, we may induce additionally to the Kerr nonlinearity a motion squeezing by imposing the vibrational resonance condition $\Phi=2\omega_{x}$. In this case, the time-averaged Hamiltonian becomes (see Appendix \ref{EH} for more details)
	\begin{figure}
		\includegraphics[width=0.45\textwidth]{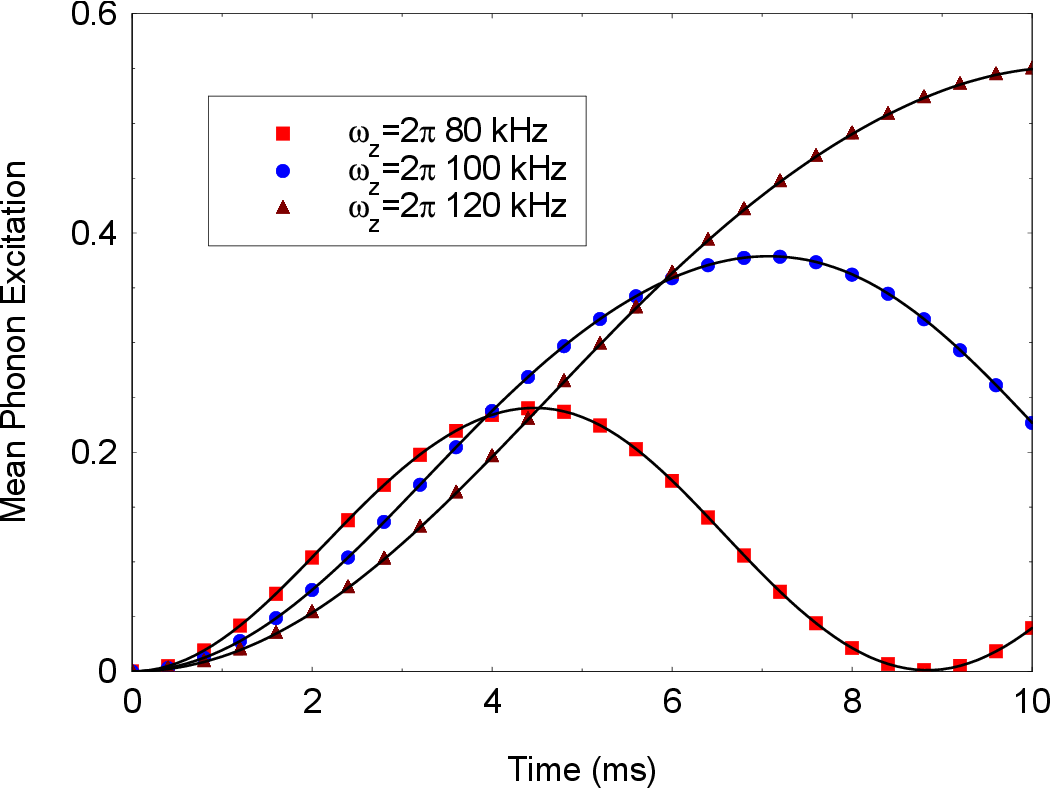}
		\caption{Mean phonon excitation of the radial vibrational mode $\bar{n}_{x}$ as a function of time for different axial trap frequencies. We compare the mean phonon excitation derived from the original Hamiltonian (\ref{Hr}) and the effective Hamiltonian (\ref{Heff}) (solid lines). The parameters are set to $l_{0}=0.05$ mm, $\omega_{x}/2\pi=1.2$\,MHz, $\Phi/2\pi=2.4$\,MHz, and $F_{z}=700$\,yN. We assume that both the axial and the radial motion modes are prepared in their ground states.}
		\label{fig2}
	\end{figure}
	\begin{equation}
		\hat{H}_{\rm eff}=-\hbar K \hat{n}^{2}_{x}-\hbar\omega \hat{n}_{x}-\hbar\epsilon (\hat{a}^{\dag 2}_{x}+\hat{a}^{2}_{x})+\hat{H}_{\rm res},\label{Heff}
	\end{equation}
	where
	\begin{eqnarray}
		&&K=\lambda^{2}\left(\frac{4}{\omega_{z}}-\frac{2\omega_{z}}{\Phi^{2}-\omega^{2}_{z}}\right),\quad \omega=\lambda^{2}\left(\frac{4\Phi-2\omega_{z}}{\Phi^{2}-\omega^{2}_{z}}+\frac{4}{\omega_{z}}\right),\notag\\
		&&\epsilon=\frac{\Omega\lambda\Phi}{\Phi^{2}-\omega^{2}_{z}}.\label{couplings}
	\end{eqnarray}
	The result implies that due to the funnel-shaped three-dimensional potential, an ion driven periodically along the axial degree of freedom experiences motion squeezing $(\hat{a}^{\dag2}_{x}+\hat{a}^{2}_{x})$ and Kerr non-linearity $\hat{n}^{2}_{x}$ of the radial ion's motion. Remarkably as can be seen from Fig.~\ref{fig1}, the nonlinear coupling $K$ and the effective phonon frequency $\omega$ can be controlled by the radial trap frequency $\omega_{x}$, the driving frequency $\Omega$ and the parameter of the  funnel-shaped trap, $l_{0}$. The squeezing rate $\epsilon$ becomes independent from $\Phi$ for $\Phi\gg\omega_{z}$ and it can be enhanced by increasing the driving frequency $\Omega$. The Kerr-nonlinearity increases with $\omega_{x}$ such that in the limit $\omega_{x}\gg\omega_{z}$, we have $K\approx r^{2}_{0z}\omega_{x}^{2}/l^{2}_{0}\omega_{z}$ and $K\approx\omega$, see Fig. \ref{fig1}(a). Finally, the residual term
	\begin{equation}
		\hat{H}_{\rm res}=-\frac{4\hbar\lambda^{2}\Phi}{\Phi^{2}-\omega^{2}_{z}}(\hat{n}_{z}+2\hat{n}_{z}\hat{n}_{x}),\label{res}
	\end{equation}
	describes phonon-phonon coupling between the axial and radial modes. The first factor in (\ref{res}) describes effective axial phonon frequency. The second term gives rise to an effective cross Kerr-nonlinearity between the axial and radial motion modes.  Since the Hamiltonian (\ref{Heff}) is diagonal in the axial Fock basis, the effect of $\hat{H}_{\rm res}$ on the dynamics of the radial mode is expressed as renormalisation of the effective radial phonon frequency. In Fig. \ref{fig2} we show the time evolution of $\langle \hat{n}_{x}(t)\rangle$ for various axial trap frequencies. As expected the effective Hamiltonian describes the exact dynamics given by Eq. (\ref{H1}) very accurately. Decreasing $\omega_{z}$ the effect of the Kerr nonlinear coupling becomes more prominent, which is in agreement with Eq. (\ref{couplings}).
	
	\subsection{Spin-Motion Coupling}
	Here we study the effect of spin-dependent light forces applied on a single ion in a tapered trap. We consider a laser-ion interaction which couples the motional mode with the internal spin states. We assume that the ion possesses two metastable electronic states, $\left|\uparrow\right\rangle$ and $\left|\downarrow\right\rangle$, with a transition frequency $\omega_{0}$. Two pairs of non-copropagating laser beams with beat note frequencies $\omega_{0}\pm\mu$ simultaneously address the spin along the axial $z$-direction \cite{Wineland1998,Haffner2008,Schneider2012} (see Appendix \ref{SDSKO} for more details). Transforming the interaction Hamiltonian into rotating frame with respect to $\hat{H}_{0}=\omega_{x}\hat{n}_{x}+\omega_{z}\hat{n}_{z}+\frac{\omega_{0}}{2}\sigma_{z}$ and assuming rotating-wave approximation and Lamb-Dicke limit, we arrive at
	\begin{eqnarray}
		\hat{H}^{\prime}_{R}&=&\hbar\lambda\{(e^{i(2\omega_{x}+\omega_{z})t}\hat{a}^{\dag 2}_{x}\hat{a}^{\dag}_{z}+{\rm h.c.})+(e^{i(2\omega_{x}-\omega_{z})t}\hat{a}^{\dag 2}_{x}\hat{a}_{z}\notag\\
		&&+{\rm h.c.})\}+\hbar\lambda(\hat{a}^{\dag}_{z}e^{i\omega_{z}t}+\hat{a}_{z}e^{-i\omega_{z}t})(2\hat{n}_{x}+1)\notag\\
		&&-i\hbar g\sin(\mu t)(\hat{a}^{\dag}_{z}e^{i\omega_{z}t}-\hat{a}_{z}e^{-i\omega_{z}t})\sigma_{x},\label{Hr1}
	\end{eqnarray}
	where $\sigma_{x,y,z}$ are the Pauli matrices and $g$ is the spin-motion coupling. The last term in (\ref{Hr1}) describes the spin-phonon coupling, provided that the beat note frequency is set to $\mu=2\omega_{x}$. Then the time-averaged Hamiltonian takes the form
	\begin{equation}
		\hat{H}^{\prime}_{\rm eff}=-\hbar K^{\prime} \hat{n}^{2}_{x}-\hbar\omega^{\prime} \hat{n}_{x}-\hbar\epsilon^{\prime} (\hat{a}^{\dag 2}_{x}+\hat{a}^{2}_{x})\sigma_{x}+\hat{H}_{\rm res},\label{Heff1}
	\end{equation}
	The couplings $K^{\prime}$, $\omega^{\prime}$, and $\epsilon^{\prime}$ in (\ref{Heff1}) are identical in shape to (\ref{couplings}) by the replacements $\Phi\rightarrow\mu$ and $\Omega\rightarrow g$, see Appendix \ref{SDSKO}.

	\section{Creation of Squeezed Kerr state}\label{CSKS}
	Control over the frequency and magnitude of the force term (\ref{force}) allows us to implement the Kerr unitary operator as well as the squeeze operator. In fact, if we set $F_{z}=0$, the unitary evolution is described by the Kerr unitary operator $\hat{U}_{\rm K}(K,\omega)=e^{iK t \hat{n}^{2}_{x}}e^{i\omega t \hat{n}_{x}}$. On the other hand, in the limit $\epsilon\gg K,\omega$ the unitary evolution is described by the squeeze operator $\hat{S}(\epsilon)=e^{i\epsilon t(\hat{a}^{\dag 2}_{x}+\hat{a}^{2}_{x})}$. Let us first consider the creation of a Kerr-coherent state, defined by $|\Psi_{K}\rangle=\hat{U}_{K}|\alpha\rangle$ where $|\alpha\rangle$ is the coherent state. Such Kerr nonlinear interaction can be used for creation of coherent cat states which are suitable for quantum error correction and quantum metrology \cite{He2023,Ofek2016}. Consider a time-varying electric field in resonance with the trap frequency being applied along the radial direction. The effect of the electric field is to displace the motion amplitude along the radial direction with the displacement operator $\hat{D}(\beta)=e^{-i\beta \hat{G}}$ and $\beta$ being the parameter we wish to estimate and $\hat{G}=(\hat{a}^{\dag}_{x}e^{i\theta}+\hat{a}_{x}e^{-i\theta})/\sqrt{2}$ being the generator of the unitary evolution. In a typical quantum metrology scheme, the system is prepared in state $|\Psi_{0}\rangle$ that subsequently evolves under the action of $\hat{D}(\beta)$ so that we have $|\Psi_{\beta}\rangle=\hat{D}(\beta)|\Psi_{0}\rangle$. The statistical uncertainty $\Delta\beta^{2}$ of the parameter estimation is given by the quantum Cramer-Rao bound $\Delta\beta^{2}\geq F^{-1}_{Q}(\beta)$, where $F_{Q}(\beta)$ is the quantum Fisher information (QFI). For pure states, QFI is $F_{Q}(\beta)=4\Delta^{2}\hat{G}$, where $\Delta^{2}\hat{G}=\langle \hat{G}^{2}\rangle-\langle \hat{G}\rangle^{2}$ is the variance of $\hat{G}$ with respect to the initial state. It is well known that for the initial coherent state $|\Psi_{0}\rangle=|\alpha\rangle$ the statistical uncertainty of the force estimation is $\Delta\beta^{2}=\frac{1}{2}$.
	\begin{figure}
		\includegraphics[width=0.45\textwidth]{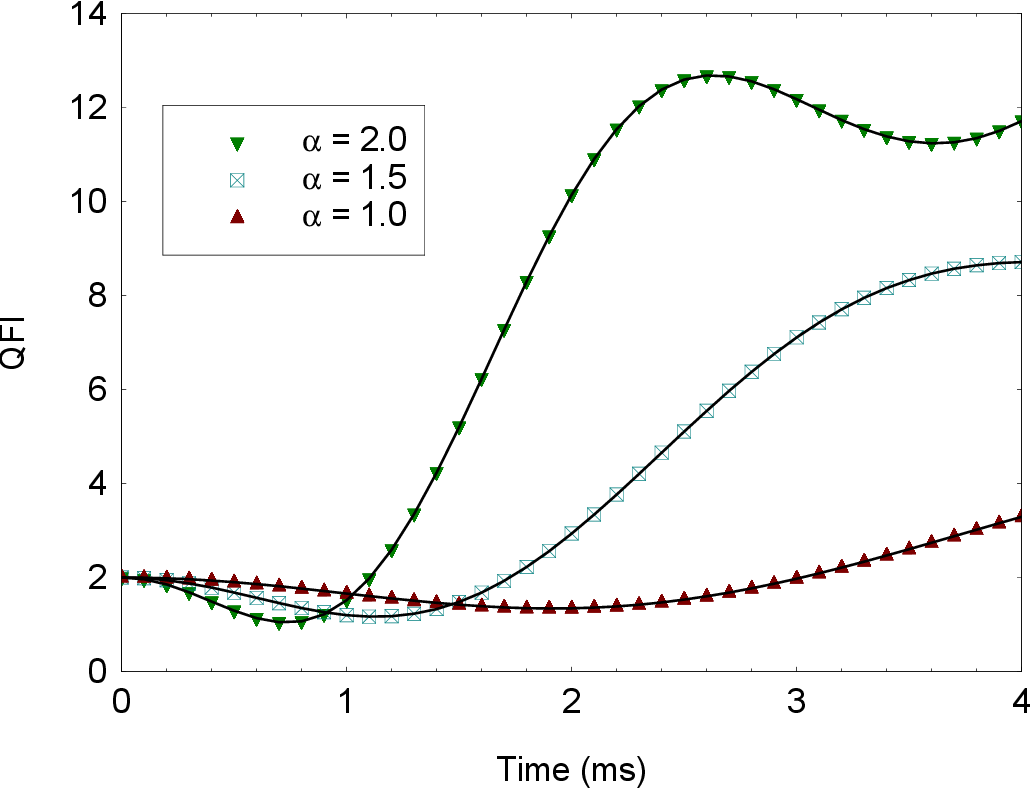}
		\caption{Time evolution of the quantum Fisher information (QFI) for various initial displacement amplitudes $\alpha$. We compare the exact result for QFI derived from the original Hamiltonian (\ref{Hr}) and the effective Hamiltonian (\ref{Heff}) (solid lines). The parameters are set to $l_{0}=0.05$ mm, $\omega_{z}/2\pi=100$ kHz, $\omega_{x}/2\pi=1.2$ MHz, $\Phi/2\pi=2.4$ MHz, and $F_{z}=0$.}
		\label{fig3}
	\end{figure}
	In Fig. \ref{fig3}, we plot the QFI assuming an initial Kerr-coherent state $|\Psi_{K}\rangle$. We compare the exact dynamics with Hamiltonian (\ref{H1}) to the effective picture described by Hamiltonian (\ref{Heff}) and observe a very good agreement between the effective approximation and the exact result. We notice that $F_{Q}(\beta)>2$ corresponds to a reduction in quantum noise below the classical limit, so the statistical uncertainty becomes $\Delta\beta^{2}<\frac{1}{2}$. For short interaction times, the QFI is $F_{Q}(\beta)<2$, which implies that there is no improvement in the measurement statistics with respect to the initial coherent state. For longer interaction times, the QFI increases above the classical limit, and the effect is more prominent for larger values of $\alpha$ (see Appendix \ref{QFI}). Recently, the metrological usefulness of the Kerr-nonlinear oscillator and the optimal measurement of displacement estimation has been discussed in \cite{Guo2024}.

	\section{Ion String in a tapered ion trap}\label{IS}
	In the following, we consider a system of $N$ ions in a tapered ion trap. The potential energy of the system consists of harmonic potential, mutual Coulomb repulsion and a position-dependent nonlinear motion coupling due to the funnel-shaped geometry:
	\begin{eqnarray}
		\hat{V}&=&\frac{m}{2}\sum_{\chi=x,y,z}\sum_{j=1}^{N}\omega^{2}_{\chi}\hat{r}^{2}_{\chi,j}+
		\frac{m\omega^{2}_{x}}{l_{0}}\sum_{j=1}^{N}\hat{r}_{z,j}\hat{r}^{2}_{x,j}
		+\frac{m\omega^{2}_{y}}{l_{0}}\sum_{j=1}^{N}\hat{r}_{z,j}\hat{r}^{2}_{y,j}\notag\\
		&&+\frac{e^{2}}{8\pi\epsilon_{0}}\sum_{j\neq k}\left\{\sum_{\chi=x,y,z}(\hat{r}_{\chi,j}-\hat{r}_{\chi,k})^{2}\right\}^{-\frac{1}{2}},\label{V}
	\end{eqnarray}
	where $\hat{\vec{r}}_{j}=(\hat{r}_{x,j},\hat{r}_{y,j},\hat{r}_{z,j})$ is the radius vector of the $j$-th ion and $\epsilon_{0}$ is the permittivity of free space. As long as the radial trap frequencies are much larger than the axial trap frequency $\omega_{x,y}\gg\omega_{z}$, the ions are arranged in a linear configuration and occupy equilibrium positions $z^{0}_{j}$ along the $z$ axis. From Eq. (\ref{V}) it follows that the equilibrium positions $z^{0}_{j}$ are not affected by the geometrically induced nonlinear terms in (\ref{V}). Hence, the position operator can be written as
	\begin{equation}
		\hat{\vec{r}}_{i}=(z^{0}_{i}+\delta \hat{r}_{z,i})\hat{e}_{z}+\delta \hat{r}_{x,i}\hat{e}_{x}+\delta \hat{r}_{y,i}\hat{e}_{y},\label{HS}
	\end{equation}
	where $\delta \hat{r}_{\alpha,i}$ are the displacement operators, subjected to the potential (\ref{V}).
	\begin{figure}
		\includegraphics[width=0.47\textwidth]{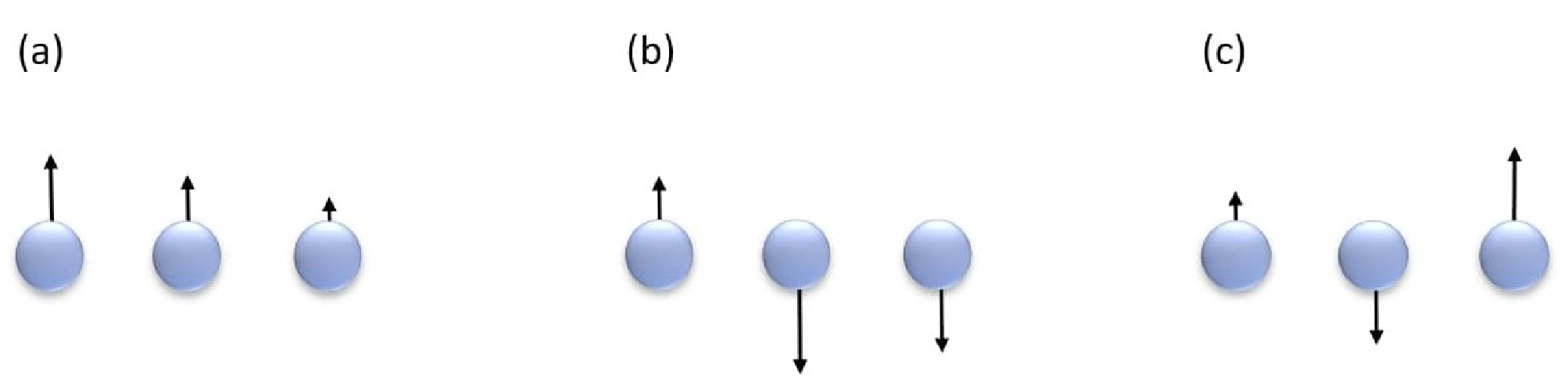}
		\caption{Example of a ion chain with three ions. Arrows qualitatively indicate the ion participation in each radial motion mode.}
		\label{fig44}
	\end{figure}

\begin{figure}
		\includegraphics[width=0.45\textwidth]{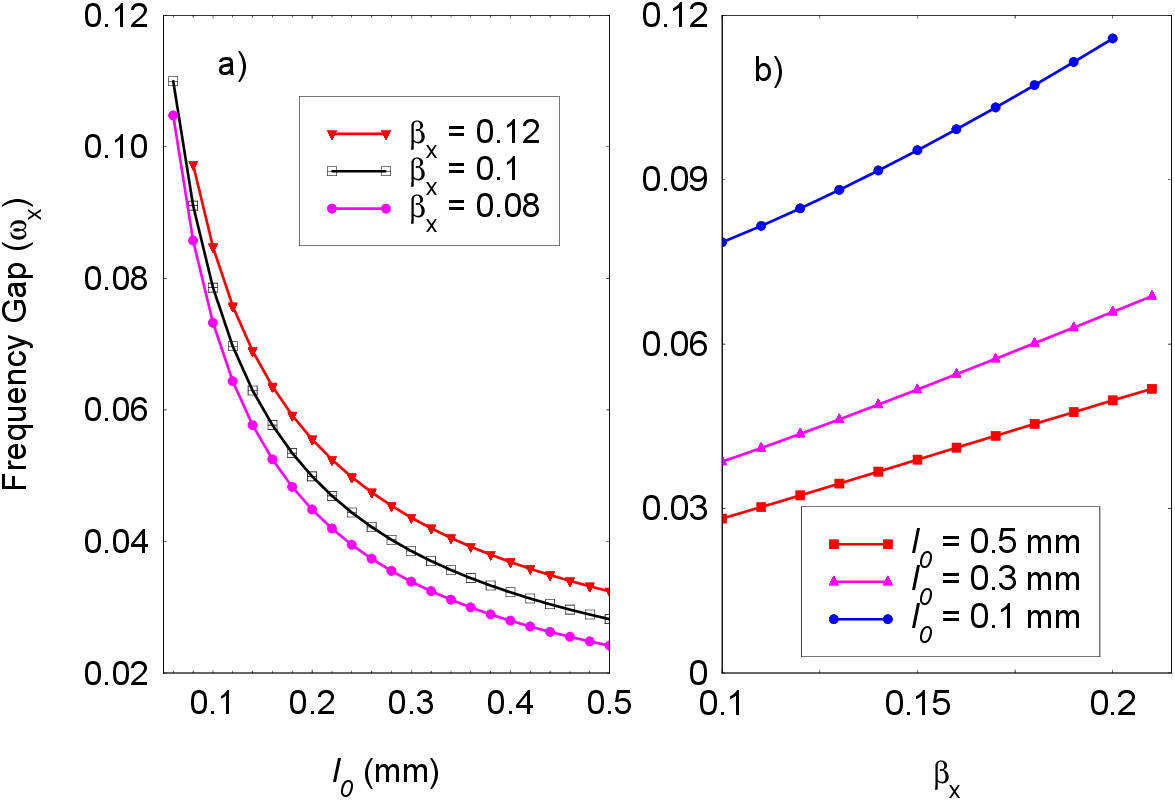}
		\caption{(a) Frequency gap between the highest center-of-mass mode frequency  and the next motion mode as a function of $l_{0}$ for various aspect ratios $\beta_{x}$. The frequency gap as a function of $\beta_{x}$ for different $l_{0}$. We assume ion string with $N=10$ ions and $l=10$ $\mu$m.}
		\label{fig4}
	\end{figure}

	After Taylor expansion of the potential $\hat{V}$, the ions' vibrations are described by the Hamiltonian
	\begin{eqnarray}
		\hat{H}_{\rm}&=&\sum_{\chi}\sum_{j}\frac{\hat{p}^{2}_{\chi,j}}{2m}+\frac{m}{2}\sum_{\chi}\sum_{j,k}\omega^{2}_{\chi}A_{jk}^{(\chi)}\delta \hat{r}_{\chi,j}\delta \hat{r}_{\chi,k}\notag\\
		&+&\!\!\frac{m\omega^{2}_{x}}{l_{0}}\sum_{j}\delta \hat{r}_{z,j}\delta \hat{r}^{2}_{x,j}+\frac{m\omega^{2}_{y}}{l_{0}}\sum_{j}\delta \hat{r}_{z,j}\delta \hat{r}^{2}_{y,j}+\hat{H}_{\rm NL}.\label{nl}
	\end{eqnarray}
	The first line in (\ref{nl}) describes the harmonic collective vibrations of the ion string. The second line in (\ref{nl}) describes the nonlinearities arising from the trap geometry as well as the nonlinear coupling $\hat{H}_{\rm NL}$ describing the higher-order motion coupling caused by Coulomb repulsion between ions \cite{Marquet2003}. Within the harmonic approximation the elements of $(N\times N)$ real and symmetric matrices $A_{jk}^{(\chi)}$ are given by
	\begin{equation}
		A_{i,j}^{(z)}=\left\{\begin{array}{c}
			1+\sum_{r=1,r\neq j}^{N}\frac{2}{|u_{j}-u_{r}|^{3}},\ (i=j), \\
			-\frac{2}{|u_{i}-u_{j}|^{3}},\ (i\neq
			j),
		\end{array}\right.
	\end{equation}
	and respectively
	\begin{equation}
		A_{i,j}^{(k)}=\left\{\begin{array}{c}
			1+\frac{2 lu^{0}_{i}}{l_{0}}-\beta^{2}_{k}\sum_{r=1,r\neq j}^{N}\frac{1}{|u_{j}-u_{r}|^{3}},\ (i=j), \\
			\frac{\beta^{2}_{k}}{|u_{i}-u_{j}|^{3}},\ (i\neq
			j),
		\end{array}\right.
	\end{equation}
	where $\beta_{k}=\omega_{z}/\omega_{k}$ for $k=x,y$. Here, $u_{i}$ are the dimensionless equilibrium positions and $l=(e^{2}/2\pi\epsilon_{0}m\omega_{z})^{1/3}$ is the length scale which characterizes the average distance between the ions. We see that the axial vibration modes are not affected by the non-linear term in (\ref{HS}). Consider now the motion along the radial direction. The collective vibration frequencies are $\omega_{\chi,p}=\omega_{\chi}\sqrt{\gamma_{\chi,p}}$ ($p=1,2,\ldots,N$) where $\gamma_{\chi,p}$ are the eigenvalues of $A_{jk}^{(\chi)}$, namely $\sum_{l=1}^{N}A^{(k)}_{ik}b^{(k)}_{l,p}=\gamma_{k,p}b^{(k)}_{i,p}$ with $b^{(k)}_{l,p}$ being the normal mode eigenvectors. Since funnel-shaped geometry breaks the radial trap symmetry, the radial vibrational modes begin to depend on the parameter $l_{0}$. Such dependencies modify the vibration spectrum of the ion string. In Fig. \ref{fig44} we show the radial motion of three ions in the tapered trap. Since the radial trap frequency varies with the axial distance, the amplitudes of the radial motion decrease with $z$ for the highest frequency center-of-mass motion. Breaking of the reflection symmetry also implies that the middle ion oscillates in the lower- frequency rocking mode. In Fig. \ref{fig4} we show the radial frequency gap between the highest center-of-mass mode frequency and the next lower motion mode frequency. Decreasing $l_{0}$ the effect of nonlinearity increases which leads to larger frequency gap; see Fig. \ref{fig4}(a). The effect is more prominent for a larger aspect ratio $\beta_{x}$, as shown in Fig. \ref{fig4}(b). Consider, for example, an ion string with $N=10$ ions and and $\beta_{x}=0.1$. For the conventional ion trap, the frequency gap is $\Delta\omega=\omega_{x}(1-\sqrt{1-\beta^{2}_{x}})\approx 5\times 10^{-3}\omega_{x}$. For a funnel-shaped ion trap with $l_{0}=1$mm, we have $\Delta\omega\approx 2\times 10^{-2}\omega_{x}$ which is approximately four times larger.
	
	\section{Summary and Outlook}\label{SO}
	In this work we have exploited the dynamics of trapped ions that arise in a funnel-shaped three-dimensional potential. The tapered geometry of the trap leads to a position- dependent radial trap frequency and, respectively, to a motion coupling between the axial and radial degrees of freedom. We have shown that the asymmetric trap geometry naturally induces Kerr nonlinearity of the radial mode, as well as a cross-Kerr coupling between the axial and radial modes for a single trapped ion, with magnitudes controlled by the trap frequencies. Furthermore, the off-resonant axial displacement causes motion squeezing of the radial mode, whose strength can be selected by modification of the amplitude of the applied electric field. This leads to a realization of the mechanically squeezed Kerr oscillator with a single ion in a tapered trap.
	
	Due to excellent control over the frequency and magnitude of the force, we are able to incorporate the squeeze operator and the unitary Kerr operator into the time evolution of our system. The strong Kerr nonlinearity can be used for the realization of non-Gaussian quantum gates which are significant parts of the universal set of continuous variable quantum gates. We have discussed the preparation of non-Gaussian quantum states that are suitable for high-precision quantum metrology.
	In addition, spin-motion coupling in a system has been discussed through the application of a spin-dependent force on a single ion in the tapered trap. The coupling between internal and motional degrees of freedom is crucial for quantum state preparation and quantum computation.
	
	Finally, we have discussed the vibrational mode spectrum of an ion chain in a taper trap. The trap geometry breaks the reflection symmetry, which modifies the ion's motion in comparison with the conventional ion trap. We have shown that the radial frequency gap appeared to be strongly related to the distance from the trap axis, and thus can serve as a reliable control parameter. Increasing the trap nonlinearity leads to a larger frequency spacing between the vibrational modes, which improves the precision in state preparation and quantum gate operations.

    The trap design is also crucial for maximizing the Kerr non-linearity. For a sensitive measurement setup, compensation in a tapered ion trap is harder to realize than in a linear ion trap due to different geometry- related field penetration of compensation electrodes. We recommend implementing the tapered Paul trap by a linear trap geometry with segmented radio-frequency electrodes in conjunction with different driving amplitudes. This has the advantage that parameters such as $l_0$ become tuneable and that micromotion compensation should be achievable for all ions along the trap axis. The presented results are crucial for designing an optimal trap configuration to maximize the Kerr nonlinearity and allowing for the implementation of high-fidelity CV quantum gates.

	\section*{Acknowledgments}
	B. S. N. and P. A. I. acknowledge the Bulgarian national plan for recovery and resilience, contract BG-RRP-2.004-0008-C01 (SUMMIT: Sofia University Marking Momentum for Innovation and Technological Transfer), project number 3.1.4. M.G. and K.S were supported by the Deutsche Forschungsgemeinschaft (DFG, German Research Foundation) -
	Projects Nos. 499241080, 384846402 - through the QuantERA grant ExTRaQT and the Research Unit Thermal Machines in the Quantum World (FOR 2724).

	\begin{appendix}
		\section{Effective Hamiltonian}\label{EH}
		The interaction picture Hamiltonian, given by
		Eq. (\ref{Hr}), with resonance condition $2\omega_x= \Phi$ implied, transforms into:
		\begin{flalign}
			\hat{H}_R(t)=&\lambda(2\hat{n}_x+1)(\hat{a}_z^\dagger e^{i\omega_z t} +\hat{a}_z e^{-i \omega_z t})  \notag \\
			&+\{e^{i(\Phi+\omega_z)t}\left(\lambda\hat{a}_x^{\dagger^2}\hat{a}_z^\dagger + \frac{\Omega}{2}\hat{a}_z^\dagger\right)+ {\rm h.c.}\} \notag\\
			&+\{e^{i(\Phi-\omega_z)t}\left(\lambda \hat{a}_x^{\dagger^2}\hat{a}_z - \frac{\Omega}{2}\hat{a}_z\right)+{\rm h.c.}\} \label{1}.
		\end{flalign}
		As long as the conditions $\omega_{z}\gg\lambda$ and $(\Phi\pm\omega_{z})\gg\lambda,\Omega/2$ are satisfied, we can perform time averaging of the rapidly oscillating terms. The effective Hamiltonian of a system can be expressed in a rather compact way by the following sum of commutators, described in \cite{James2007}:
		\begin{equation}
			\hat{H}_{\rm eff}=\sum_{n,m=1} \frac{1}{\hbar\bar\omega_{mn}} [\hat{h}_m^\dagger, \hat{h}_n] e^{i(\omega_m - \omega_n)t},\label{2}
		\end{equation}
		with  $\hat{h}_{m,n}$ being the harmonic terms constructing the interaction Hamiltonian and $\bar\omega_{mn}$ are the harmonic averages of $\omega_m$ and ${\omega_n}$, where $\frac{1}{\bar\omega_{mn}} = \frac{1}{2}\left(\frac{1}{\omega_m} - \frac{1}{\omega_n}\right)$. Using Eqs. (\ref{1}) and (\ref{2}) we obtain the effective Hamiltonian (\ref{Heff}) in the main text.

        \section{Spin-dependent squeezed Kerr oscillator}\label{SDSKO}

        We assume that the trapped ion interacts simultaneously with bichromatic laser fields with laser frequencies $\omega^{r}_{L}=\omega_{0}-\mu$, $\omega^{b}_{L}=\omega_{0}+\mu$ and laser phases $\phi_{r}$, $\phi_{b}$ along the trap axis. The interaction Hamiltonian in the Lamb-Dicke limit which describes the ion-laser interaction is $\hat{H}_{I}=\Omega\eta\sin(\mu t)\sigma(\phi_{+})(\hat{a}_{z}^{\dag}e^{i\phi_{-}}+\hat{a}_{z}e^{-i\phi_{-}})$, where $\sigma(\phi_{+})=(\sigma_{+}e^{i\phi_{+}}+\sigma_{-}e^{-i\phi_{+}})$ \cite{Lee2005}. Here $\eta$ is the Lamb-Dicke parameter and $\phi_{\pm}=(\phi_{b}\pm\phi_{r})/2$. Transforming the Hamiltonian into rotating frame with respect to $\hat{H}_{0}$ and set the resonance condition $\mu=2\omega_{x}$ we obtain
       \begin{flalign}
			\hat{H}_R(t)=&\lambda(2\hat{n}_x+1)(\hat{a}_z^\dagger e^{i\omega_z t} +\hat{a}_z e^{-i \omega_z t})  \notag \\
			&+\{e^{i(\mu+\omega_z)t}\left(\lambda\hat{a}_x^{\dagger^2}\hat{a}_z^\dagger + \frac{g}{2}\hat{a}_z^\dagger\sigma_{x}\right)+ {\rm h.c.}\} \notag\\
			&+\{e^{i(\mu-\omega_z)t}\left(\lambda \hat{a}_x^{\dagger^2}\hat{a}_z - \frac{g}{2}\hat{a}_z\sigma_{x}\right)+{\rm h.c.}\} \label{1}.
		\end{flalign}
Here $g=\eta\Omega$ is the spin-motion coupling and we set $\phi_{+}=0$ and $\phi_{-}=\pi/2$. The time-averaged dynamics is given by
\begin{equation}
		\hat{H}^{\prime}_{\rm eff}=-\hbar K^{\prime} \hat{n}^{2}_{x}-\hbar\omega^{\prime} \hat{n}_{x}-\hbar\epsilon^{\prime} (\hat{a}^{\dag 2}_{x}+\hat{a}^{2}_{x})\sigma_{x}+\hat{H}_{\rm res},
	\end{equation}
where
\begin{eqnarray}
		&&K^{\prime}=\lambda^{2}\left(\frac{4}{\omega_{z}}-\frac{2\omega_{z}}{\mu^{2}-\omega^{2}_{z}}\right),\quad \omega^{\prime}=\lambda^{2}\left(\frac{4\mu-2\omega_{z}}{\mu^{2}-\omega^{2}_{z}}+\frac{4}{\omega_{z}}\right),\notag\\
		&&\epsilon^{\prime}=\frac{g\lambda\mu}{\mu^{2}-\omega^{2}_{z}}.
	\end{eqnarray}
        Hence the spin-dependent laser force along the trap axis creates a spin-dependent squeezing of the radial mode.
	\section{Quantum Fisher Information for a coherent Kerr state}\label{QFI}
		Here we derive an expression for the QFI assuming that the system is prepared initially in the coherent Kerr state. We can rewrite the unitary operator as $\hat{U}_{\rm K}(K,\omega)=e^{iK t \hat{n}_{x}(\hat{n}_{x}-1)}e^{i(K+\omega )t \hat{n}_{x}}$. Then we have \cite{Gerry1994}
		\begin{eqnarray}
			&&\langle \hat{a}\rangle=e^{i(K+\omega)t}\alpha e^{|\alpha|^{2}(e^{2i K t}-1)},\notag\\
			&&\langle \hat{a}^{2}\rangle=e^{2i(K+\omega)t}\alpha^{2} e^{2i K t}e^{|\alpha|^{2}(e^{4i K t}-1)}.\label{a}
		\end{eqnarray}
		The QFI is $F_{\rm Q}(\beta)=4(\langle \hat{G}^{2}\rangle-\langle \hat{G}\rangle^{2})$ where the average is performed with respect to $|\Psi_{K}\rangle=\hat{U}_{K}|\alpha\rangle$. Using (\ref{a}) the QFI is given by
		\begin{eqnarray}
			F_{Q}(\beta)&=&2\{(e^{2i(K+\omega)t}\alpha^{2} e^{2i K t}e^{|\alpha|^{2}(e^{4i K t}-1)}+{\rm h.c.})+2|\alpha|^{2}+1\notag\\
			&&-(e^{i(K+\omega)t}\alpha e^{|\alpha|^{2}(e^{2i K t}-1)}+{\rm h.c.})^{2}\}. \label{QFI}
		\end{eqnarray}
		The expression (\ref{QFI}) analytically describes the results presented in Fig. \ref{fig3}.
		
	\end{appendix}

\end{document}